\documentclass[twocolumn,apj]{emulateapj}
\usepackage{relsize,cancel, graphicx, dcolumn,graphicx,color,booktabs,bbold, microtype,braket,mathrsfs,mathtools,simplewick,tikz,empheq,amssymb,amsmath,bm}
\usepackage[colorlinks,plainpages=false,linkcolor=black,urlcolor=blue,citecolor=blue,pdfpagemode=UseNone,pdfstartview=FitH]
{hyperref}
\usepackage[T1]{fontenc}
\usepackage{dblfloatfix}
\usepackage{fixltx2e}
\usepackage{braket}

\usepackage{amsmath}
\definecolor{MyDarkBlue}{rgb}{0, 0.4,0.6}

\def \wic [#1]{\textcolor[rgb]{0,0.8,0.1}{#1}}

\begin{document}
\title{Identifying typical \ion{Mg}{2} flare spectra using machine learning}

\author{Brandon Panos\altaffilmark{1,2}, Lucia Kleint\altaffilmark{1,3}, Cedric Huwyler\altaffilmark{1}, S\"{a}m Krucker\altaffilmark{1,4}, Martin Melchior\altaffilmark{1}, Denis Ullmann\altaffilmark{2}, Sviatoslav Voloshynovskiy\altaffilmark{2}}
\
\altaffiltext{1}{University of Applied Sciences and Arts Northwestern Switzerland, Bahnhofstrasse 6, 5210 Windisch, Switzerland}
\altaffiltext{2}{University of Geneva, CUI-SIP, 1205 Geneva, Switzerland}
\altaffiltext{3}{Kiepenheuer Institut f\"ur Sonnenphysik (KIS), Sch\"oneckstrasse 6, D-79104 Freiburg, Germany}
\altaffiltext{4}{Space Sciences Laboratory, University of California, 7 Gauss Way, Berkeley, CA  94720, USA}
\begin{abstract}
IRIS performs solar observations over a large range of atmospheric heights, including the chromosphere where the majority of flare energy is dissipated. The strong Mg II h\&k spectral lines are capable of providing excellent atmospheric diagnostics, but have not been fully utilized for flaring atmospheres. We aim to investigate whether the physics of the chromosphere is identical for all flare observations by analyzing if there are certain spectra that occur in all flares. To achieve this, we automatically analyze hundreds of thousands of Mg II h\&k line profiles from a set of 33 flares, and use a machine learning technique which we call supervised hierarchical k-means, to cluster all profile shapes. We identify a single peaked Mg II profile, in contrast to the double-peaked quiet Sun profiles, appearing in every flare. Additionally, we find extremely broad profiles with characteristic blue shifted central reversals appearing at the front of fast-moving flare ribbons. These profiles occur during the impulsive phase of the flare, and we present results of their temporal and spatial correlation with non-thermal hard X-ray signatures, suggesting that flare-accelerated electrons play an important role in the formation of these profiles. The ratio of the integrated Mg II h\&k lines can also serve as an opacity diagnostic, and we find higher opacities during each flare maximum. Our study shows that machine learning is a powerful tool for large scale statistical solar analyses.
\end{abstract}
\section{Introduction}

The Interface Region Imaging Spectrograph \citep[IRIS,][]{IRIS} routinely observes flares, yet statistical analyses of an ensemble of flares are rare. IRIS observes two of the brightest chromospheric lines, the optically thick \ion{Mg}{2} h\&k resonant lines with core vacuum wavelengths at  $2803.52$ and $2796.34$ \AA. The h\&k lines sample the entire chromosphere and provide excellent quiet Sun diagnostics \citep{Paper1}, however, there exist few diagnostics based on these lines for the flaring Sun. Our goal is to use machine learning to statistically analyze several dozen flares, and answer the question as to whether there are typical spectra that would indicate similar chromospheric physics in all flares.

In the standard model, a flare is caused by the impulsive reconfiguration of the coronal magnetic fields called reconnection. This process releases on average $10^{32}$ erg/s of magnetically stored energy within a few minutes \citep[e.g.][]{FlareEnergy}. The energy is used to accelerate electrons and protons into space as well as into the thick target of the chromosphere, resulting in heating and a subsequent emission over a large band of frequencies. During this process the \ion{Mg}{2} h\&k lines differ from the usual quiet Sun profiles in a few significant ways. The central reversals and subordinate lines often go into emission, the line wings undergo substantial broadening and highly asymmetric profiles can be observed. Many attempts have been made to understand the behavior of the h\&k lines during a flare (\cite{at1}; \cite{at2}; \cite{at3}; \cite{at4}; \cite{at5}; \cite{at6}; \cite{Adam}; \cite{at7}). Recent parameter studies by \cite{Lucia} using an artificial flaring atmosphere simulated with RADYN and the partial redistribution and non-LTE radiative transfer code RH, along with single flare observations  by \cite{Kerr} have begun to extract \ion{Mg}{2} diagnostics for the flaring Sun. An important initial step is to identify all possible types of profiles that can be generated during a solar flare.

The arrival of large ground-based telescopes such as the Daniel K. Inouye Solar Telescope (DKIST) with an estimated data output of 5 PB per year \citep{DKIST} as well as the already sizeable cumulative solar databases of IRIS and SDO places the discipline of heliophysics firmly within the territory of big data. The important aspects of this data can no longer be extracted and analyzed without the use of smart algorithms. Despite this wealth of solar data, large scale statistical studies with an ensemble of flares are a rarity in the heliophysics community, while publications based on single flare observations are the norm. 

In this paper, we analyze hundreds of thousands of \ion{Mg}{2} spectral line profiles taken from 33 flares, and use a clustering algorithm to identify structures within the data set. The k-means clustering algorithm of \cite{macqueen1967} has been used in the past by \cite{0004-637X-663-2-1386} and \cite{refId0} to cluster the different shapes of  Stokes $V$ profiles in order to investigate the magnetism of the quiet Sun. In the same spirit, we will use the k-means algorithm in combination with a manual merging and splitting of clusters based on physical
relevance and in-group variance to identify the observed shapes of \ion{Mg}{2} k-line profiles produced during flares.

\begin{figure}[!tbh] 
\centering
\includegraphics[trim={0cm .8cm 0cm 0cm},clip, width=0.5\textwidth]{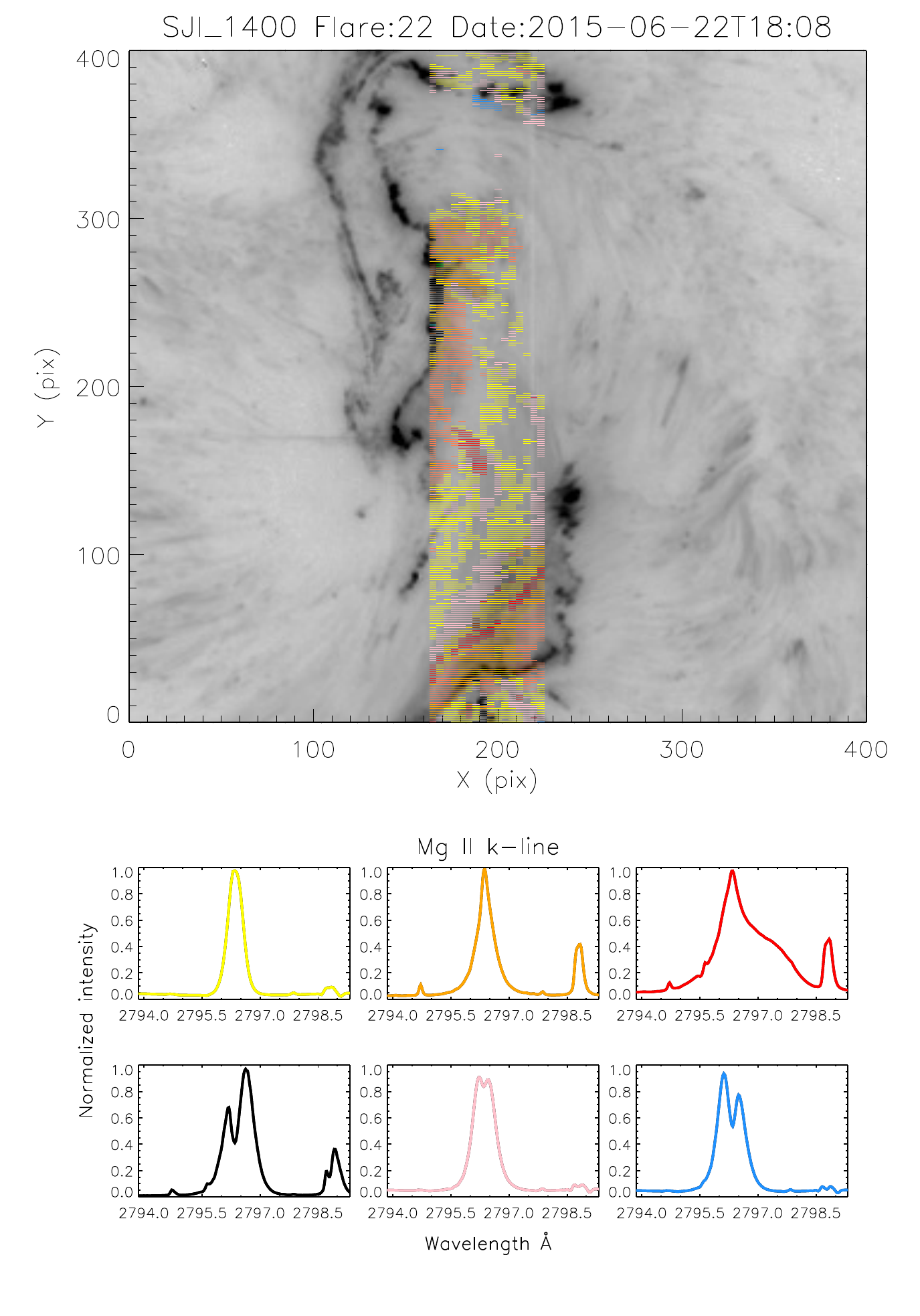}
\caption{Example of a k-means clustering for the M6.5 flare on June 6, 2015. The \ion{Mg}{2} k-line profiles for a single raster are assigned to the groups in the bottom panel that they are most similar to. The quiet Sun was not colored because its profiles belong to different groups. For the full temporal evolution of this flare, see the \textcolor{blue}{online movie}.}
\label{Intro}
\end{figure}
\section{Data and Machine learning}
\label{obs}

\subsection{IRIS}
The data for this study were taken by NASA's small explorer satellite IRIS, which was launched in 2013 and has since observed more than 400 flares according to a manually-kept list on its mission website. IRIS can take  high quality slit-jaw images (SJI) of the solar atmosphere with a maximum field of view of $175\times175~\text{arcsec}^2$ in four different passbands, covering a range of heights from the photosphere to the transition region. It is also equipped with a spectrograph that can run simultaneously at high spatial (0.33-0.4 arcsec), spectral (0.056~\AA ~in the NUV) and temporal (2s) resolutions. Various observing modes can be selected, from sit-and-stare, where the spectrograph slit remains stationary with respect to the Sun, to rasters with varying numbers of steps and slit orientations.

\subsection{K-Means}

We use an unsupervised clustering algorithm known as k-means to identify the different \ion{Mg}{2} k profiles that occur during a flare. Clustering algorithms partition similar observations into groups. If each observation consists of two features, they can be mapped as points on the cartesian plane. k-means uses additional points called centroids to group the observations. The centroids are placed on the plane and each observation is assigned to one of the centroids with a straight line. The k-means objective is to find the assignments and position of centroids that minimize the sum of all the squared Euclidian distances. This quantity is known as the "within cluster distance" $\mathcal{L}$, and is given by 
\begin{equation}
\mathcal{L} = \sum^n_{i=1}\sum^k_{j=1}\delta_{c_i,j}||x_i-\mu_j||^2.
\end{equation}
Here, $\delta_{c_i,j}$ is the Kronecker delta and $c_i$ is a label assigning each observation $x_i\in\mathbb{R}^{d_x}$ to one of the $k$ groups, so that 
\begin{equation}
\delta_{c_i,j} =
    \begin{cases}
            1, & \text{if } x_i ~\text{belongs to group}~j,\\
            0, & \text{if } x_i ~\text{does not belong to group}~j,
    \end{cases}
\end{equation}
while $\mu_j\in\mathbb{R}^{d_x}$ is the centroid of that group. To initialize the algorithm, the centroids are positioned in the locations of $k$ randomly selected data points. The k-means algorithm then minimizes the within cluster distance by iterating through a two step procedure known as coordinate descent. Firstly, the data is partitioned into groups by labeling each observation by the centroid it appears closest to 
\begin{equation}
c_i=\underset{1\leq j\leq k}{\text{arg}~\text{min}}||x_i-\mu_j||^2.
\end{equation}
Secondly, the centroids are moved to the mean position of each group according to 
\begin{equation}
\mu_j=\frac{1}{n_j}\sum^n_{i=1}\delta_{c_i,j}~x_i,
\end{equation}
where $n_j$ is the number of all observations in group $j$. Because the centroids have shifted, new labels must be assigned to all the observations in accordance to step one. This process continues until the centroids converge. The final clustering depends on the initialization of the centroids. It is customary to repeat the clustering with a number of different centroid initializations and select the clustering with the lowest cost $\mathcal{L}$. k-means was chosen for its simplicity, scalability and linear time complexity \citep{kmeans_time}. The above algorithm is easily adapted to the purposes of grouping line profiles. Instead of two features being mapped to a  2-dimensional cartesian plane, we composed each profile out of 216 $\lambda$ points and mapped them to a 216-dimensional space, i.e., $d_x=216$. k-means requires the number of clusters $k$ to be chosen by hand. There are many methods such as the "elbow technique" and silhouette analysis \citep{Silhouette}, which indicate the natural number of partitions of a data set, however, this number in many cases is left to the discretion of a professional and should not be automated. In conclusion, k-means generates $k$ groups, with each group containing a number of spectra that share similar features. These groups each have a representative spectral profile, which is the mean profile of that group called the centroid.

We experimented with a number of alternative distance metrics, most of which returned similar if not identical results to the Euclidean distance, while other distance measures found clusters that were hard to reconcile with any physical interpretation. Additionally, when updating the centroids one could select the median or a true representative point instead of the arithmetic mean. This could make the algorithm less susceptible to the effects of outliers at the price of being more computationally expensive. However, such effects are negligible if there are a sufficient number of normal data points. Furthermore, we acknowledge that there may exist a transformation and metric pair that is more suited to the clustering of spectral data, but did not wish to interject any biases into our data set.

Figure \ref{Intro} shows the result of k-means applied to a single raster of the M6.5 class flare on June 6, 2015. The bottom panel shows 6 of the 53 centroids found by our adapted k-means algorithm explained in section 2.4. The color-coded positions of the spectra for this raster have been overlaid onto the SJI. The \ion{Mg}{2} k profiles emerging from each location are assigned to one of the 53 groups based on which centroid they are most similar to.

\subsection{Data reduction}
We analyzed 26 M- and 7 X-class flares with observations in the \ion{Mg}{2} spectral window and slits positioned directly over the flaring region. The details of each flare can be found in Table \ref{Flares_List}, which includes a variety of observational modes.

Before using the k-means algorithm, the data were prepared in several ways. The \ion{Mg}{2} spectra were selected over a time interval 15 minutes before to 15 minutes after each flare, if available. The start and end times were determined from the GOES flare database.
The spectra were then cropped to a $5 ~\text{\AA}$ window which contained both the \ion{Mg}{2} k and subordinate lines consisting of the two strongest red wing $3\text{d}^2\text{D}\to3\text{p}^2\text{P}$ transitions with vacuum wavelengths at 2798.75 and 2798.82 \AA. This step not only reduces computational demands but also helps negate dimensionality problems common amongst machine learning algorithms with distance based metrics. The profiles were then interpolated using a spectral sampling of 0.025 \AA/pixel, resulting in all \ion{Mg}{2} k-line profiles having a total of 216 $\lambda$ points. Since the line intensities in data numbers (DN) can vary over two orders of magnitude, each profile was divided by its maximum intensity. This ensures that the classification is only based on the shape of the profile and not on its intensity. Data loss from incomplete spectra was automatically handled  during the cropping phase.  To visualize the results, we projected the assigned spectra onto the 1400 \text{\AA} slit-jaw image, which is sensitive to plasma at chromospheric and transition like temperatures, however, if this filter was not used for that particular observation, the 2796 \text{\AA} slit-jaw image, which sees the lower chromosphere was substituted.

\subsection{Description of the k-means pipeline}
The k-means algorithm was applied to \ion{Mg}{2} spectra collected from the subset of 4 M- and 4 X-class flares with the number of groups set to $k=80$. Flaring groups were generated by manually selecting only the spectra that appeared over the flaring regions. The algorithm was repeated 10 times with different centroid initializations and the clustering with the lowest cost $\mathcal{L}$ was selected (see section 2.2). If two or more centroids were similar, only a single centroid was retained. 

Because every profile has to be assigned to a group, it is necessary to collect a number of non-flaring centroids to filter out the non-flaring spectra. Following the same procedure above, we generated quiet Sun and sunspot groups by manually selecting only the spectra that appeared over quiet Sun and sunspot regions within the 8 selected flares. Centroids associated with small energetic events due to flux emergence were generated by running k-means over rising flux regions. This completed the unsupervised part of the algorithm.

\begin{figure}[tb] 
\centering
\includegraphics[width=0.45\textwidth]{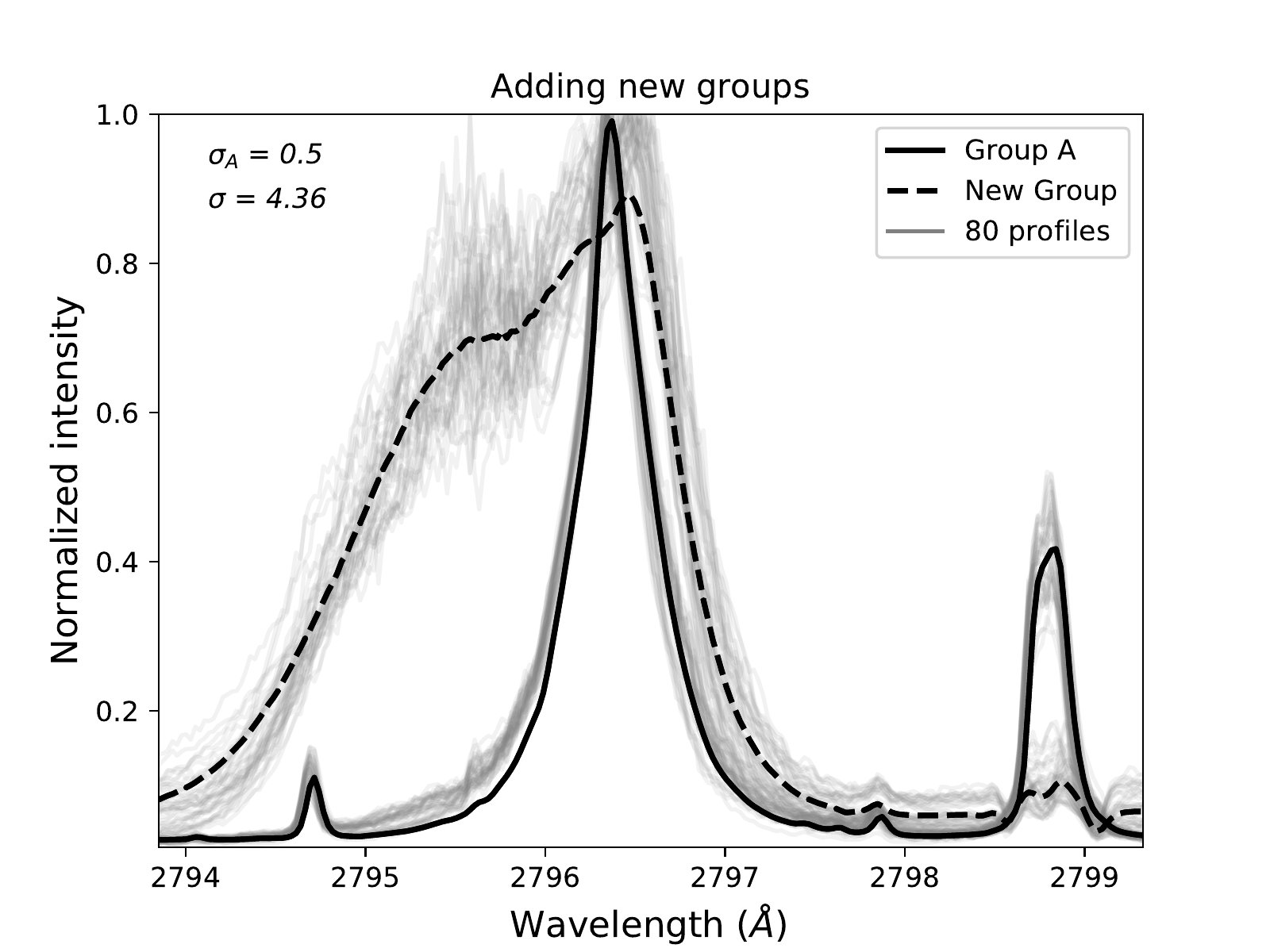}
\caption{Plot showing 80 profiles (light grey solid lines) that have been assigned to a hypothetical group A (black solid line). The variance of the profiles close to group A is given by $\sigma_{\text{A}}$, while the total variance for every profile is given by $\sigma$. A high variance indicates a potential flaring group that had not been found by the original k-means algorithm and was manually added later (broken black-line).}
\label{Adding_new_groups}
\end{figure}

\begin{figure*}[tbh] 
\centering
\includegraphics[trim={.5cm .8cm .5cm .5cm}, clip, width=.96\textwidth]{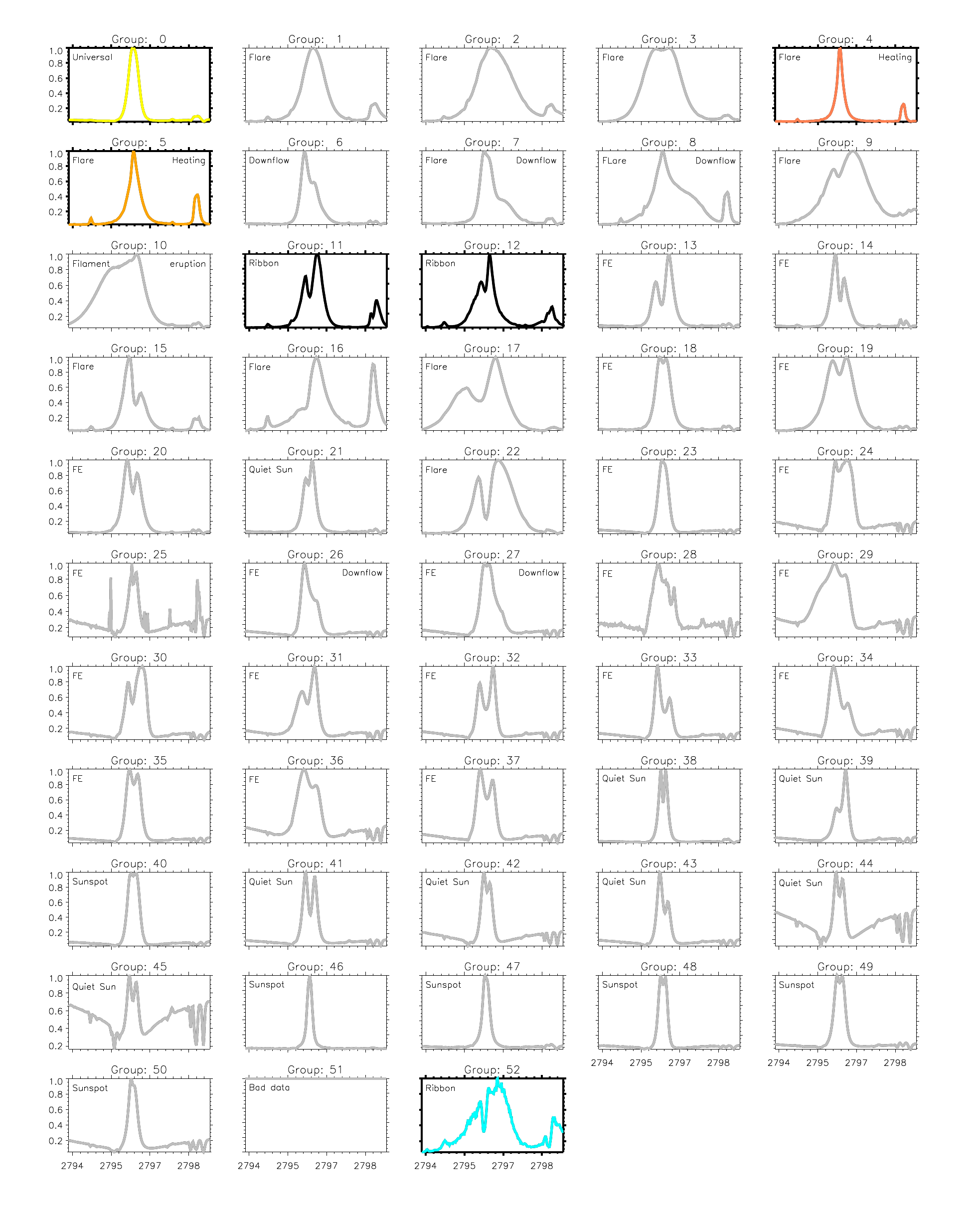}
\caption{Centroids found using the SHK algorithm, with manual merging and splitting of groups. The y-axis is in units of normalized intensity and the x-axis is in wavelengths (\AA). The groups that we analyzed are color coded and have bold borders. The event that each group is associated with appears in the top left and right corners of each panel, with FE standing for flux emergence. It is important to note that groups other than those forming the basis of this study may occur in more than one category, and were assigned descriptions based on the locations from which they were collected.}
\label{centroids}
\end{figure*}

For the remaining flares, the centroids from the clustering were passed to another algorithm (one-nearest neighbor classifier), where each profile was labeled by the centroid they appeared nearest to. Once the profiles of every flare were assigned to their groups, the variances of the flaring groups were monitored. If the variance was to high, an additional flaring group was introduced manually, with the proviso that the new group consistently corresponded with some feature on the SJI projections (see Figure \ref{Intro}). This procedure is demonstrated in Figure \ref{Adding_new_groups}, and was continued until we were satisfied that all interesting flare groups were included. The final 53 groups can be seen in Figure \ref{centroids}. Our method of clustering deviates from the original k-means algorithm in that we manually merge and split groups based on the supervision of both the variance and the SJI projections. It is unclear whether the same results could be achieved simply by increasing the initial number of centroids in the original k-means algorithm. We refer to this new clustering method as "supervised hierarchical k-means" or SHK for short.


\begin{figure}[tbh] 
\centering
\includegraphics[width=0.45\textwidth]{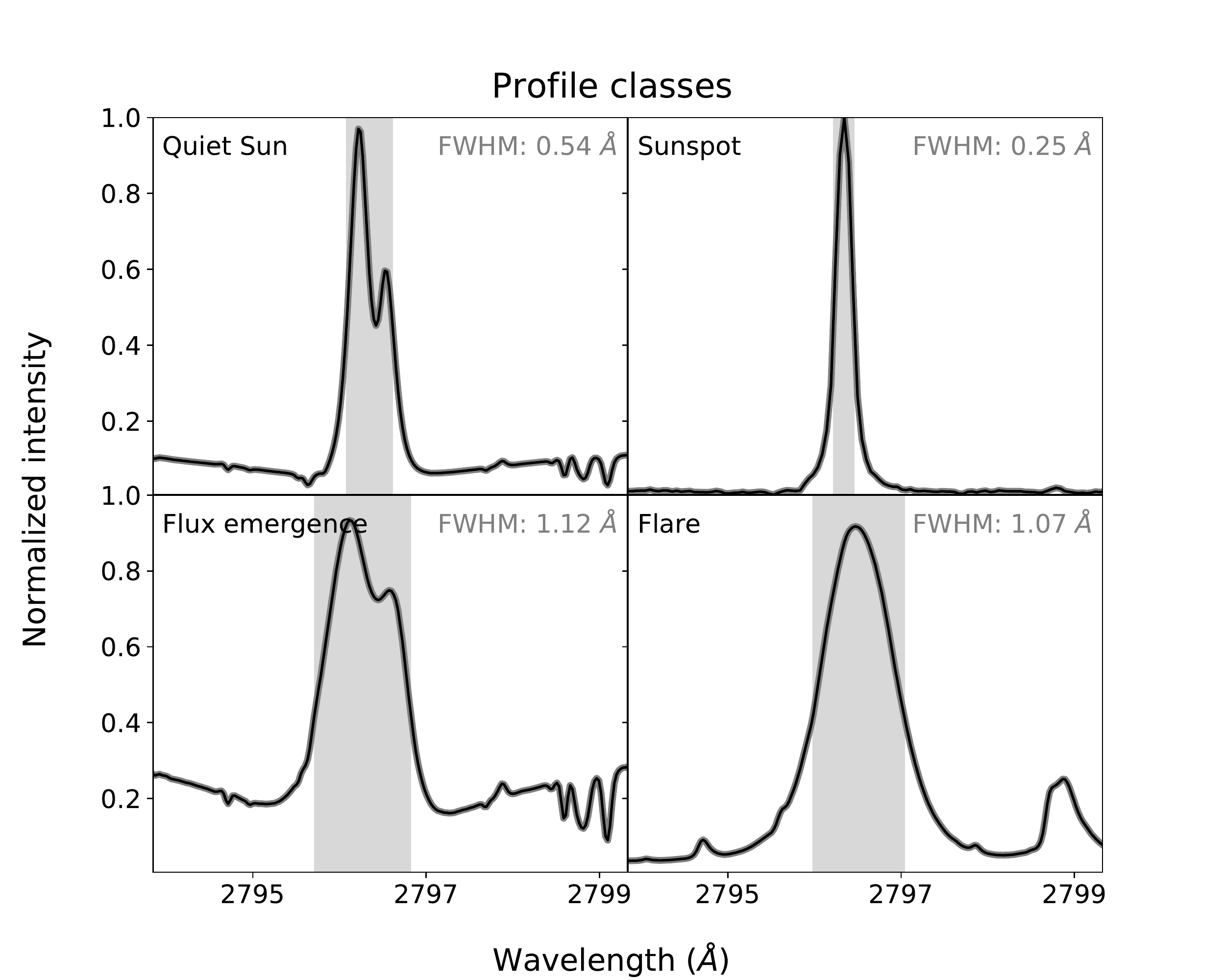}
\caption{Plot showing examples of the four profile classes. The grey shading indicates the FWHM of each centroid.}
\label{Profile_classes}
\end{figure}

\begin{figure} 
\centering
\includegraphics[width=0.5\textwidth]{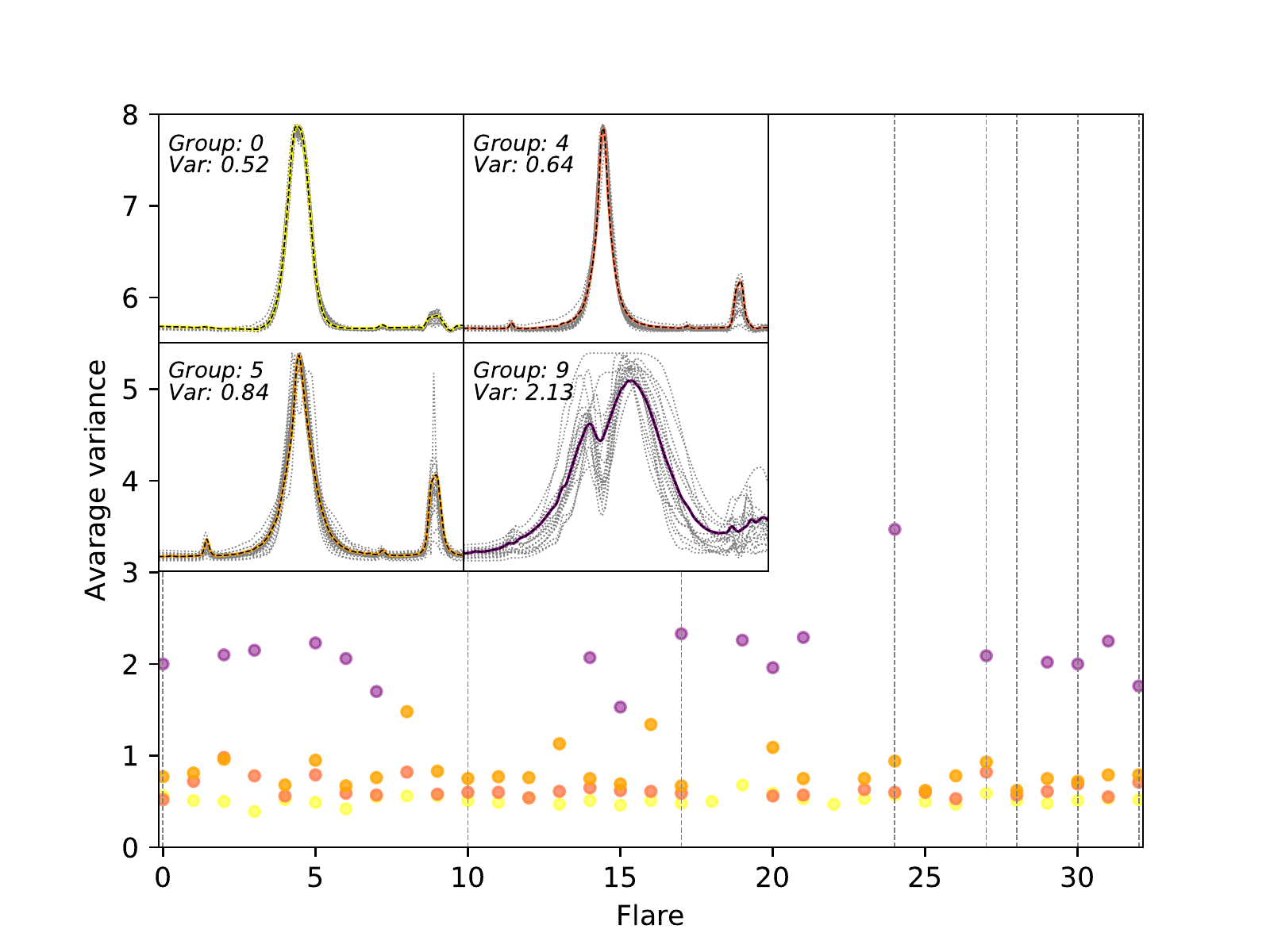}
\caption{Average variances of groups 0, 4, 5 and 9 across all 33 flares. Vertical lines indicate the flares used to train each centroid (group 9 was included manually). The inserts display group centroids with each flares average profile overplotted in grey. Var is the average variance across all flares. Group 9 has been included as an example of a group with high variance. Profiles from group 0 occur in every flare with extremely low variance and are therefore a universal flare feature.} 
\label{variances}
\end{figure}

\subsection{Centroids}
The 53 profile types fall into four main classes: Quiet Sun, sunspot, flux emergence, and flaring centroids. Figure \ref{Profile_classes} shows examples of each of the four classes. The quiet Sun profiles have characteristic central reversals, well defined blue (2kv) and red (2kr) peaks and line wings that follow the temperature structure of the lower atmosphere. Flaring profiles on the other hand can be extremely broad and are often observed without central reversals and with the subordinate lines in emission. Similarly, sunspot profiles often have a single peak, but can be distinguished from flaring profiles based on their narrow width $\sim0.25~\text{\AA}$ and lack of subordinate line emission. Flux emergence profiles share many features of quiet Sun and flaring profiles. They have raised wings and are broad, however there is no subordinate line emission.

\section{Analysis}\label{analysis}
Groups 0, 4, 5, 11, 12 and 52 in Figure \ref{centroids} appear with consistent behavior throughout our data set, indicating that they are related to flares. We now analyze our findings and discuss the behavior and defining features of each of these groups. We focus on two main results: 1) There are typical flare profiles that appear in every single flare and 2) There are special flare profiles at the front of flare ribbons.

\subsection{Are there typical flare profiles?}
 
We investigate if all flares share common profiles, which would indicate that the physics of the lower solar atmosphere may be similar in all flares.

The single peaked profiles can be divided into two different types. The first is represented by centroid 0 and has small subordinate line emission and a FWHM of 0.5 \AA. The second represented by centroid 4 and 5 has a narrower convex shape with large subordinate line emission and a FWHM between 0.29 - 0.49 \AA. Broader profiles such as those belonging to group 1, 2 and 3 are rarer, have less predictable behavior and often appear with small central reversals. In Figure \ref{variances}, we have plotted the average variances of groups 0, 4, 5 and 9 for each of the 33 flares. The single peaked profiles belonging to group 0 appear in every flare with a total average variance of 0.52, and are always located over the ribbon or in regions of small brightenings. These profiles are not exclusive to solar flares and can be stimulated by an assortment of sub-flare energetic events such as local heating from small scale reconnections over flux emergence regions. We tested their prevalence by analyzing 8 non-flaring energetic events, and 1 true quiet Sun observation. They did not appear in the quiet Sun observation but were seen in 7/8 of the energetic events (see table \ref{Non_flares_List}). The profiles therefore link both high and low energetic solar activity with temporal occurrences before, after, and during the flare, and can be identified as a universal flaring profile.

Single peaked profiles from group 4 and 5 with large triplet emission occur in the wake of the ribbon front and can have long characteristic life times of $\sim1$ hour (for example flare 31 in our list). The large triplet emission associated with these groups may be due to the heating of both the upper and lower atmosphere by non-thermal electrons \citep{triplet_diagnostics}. In every flare, the frequency of these profiles is strongly correlated with the GOES 1-8 $\text{\AA}$ channel.

\begin{figure}[tb] 
\centering
\includegraphics[width=0.5\textwidth]{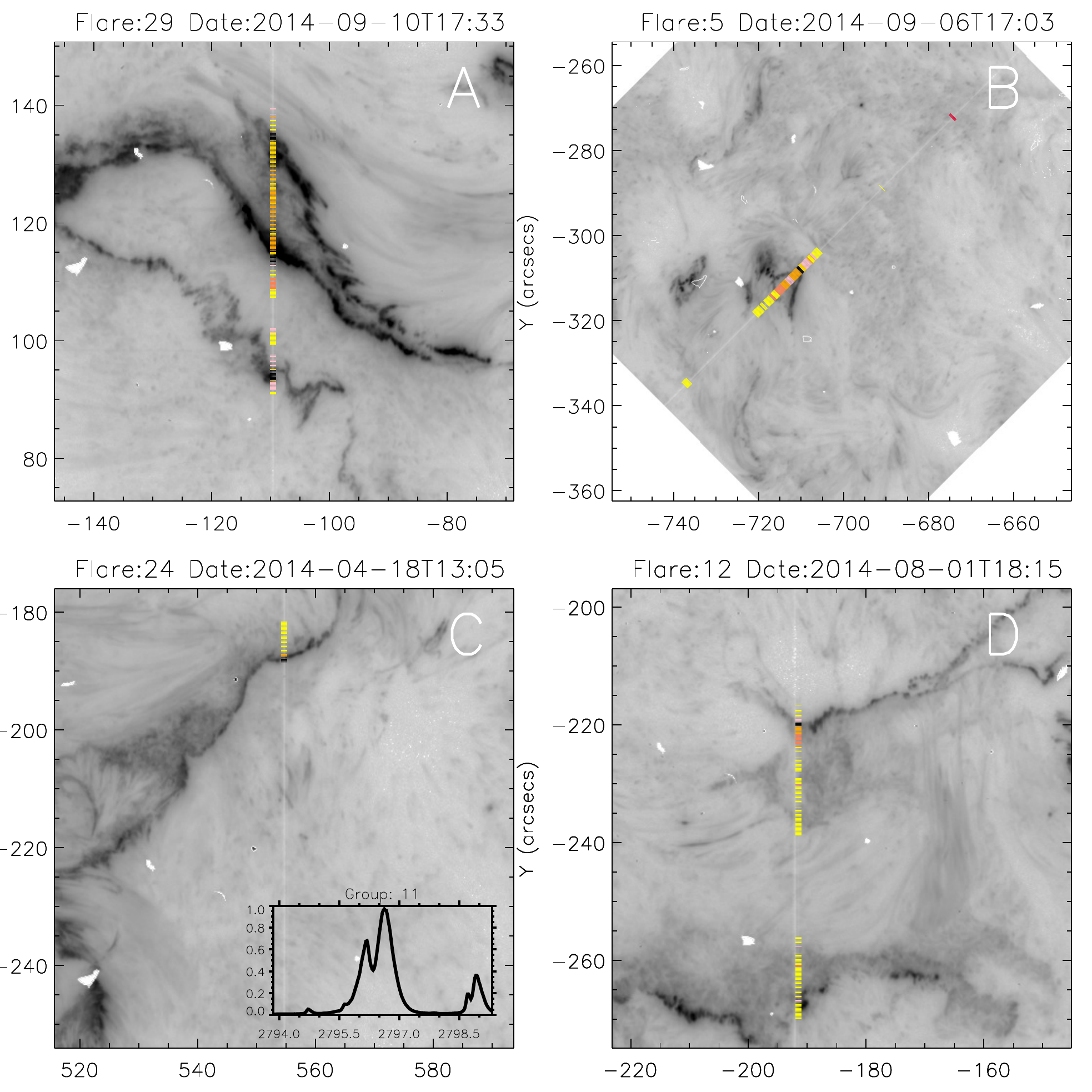}
\caption{SJIs of four different flares in the 1400 $\text{\AA}$ channel with color coded group assignments overplotted (see Figure \ref{centroids}) and x- and y-axes given in arcseconds. Profiles assigned to group 11 (bottom right of panel C) and 12 appear at the leading edge of the flare ribbons in black. The temporal evolution of each flare can be seen in the \textcolor{blue}{online movies}.}
\label{simp4}
\end{figure}

\subsection{Ribbon-front IRIS profiles} 
Here we investigate which profiles occur at ribbon fronts, where accelerated electrons are thought to reach lower atmospheric layers.
We find that profiles belonging to groups 11 and 12 occur at the leading edge of fast-moving flare ribbons, however, there are also a number of false positives generated by upflowing material. These profiles are similar to the ribbon-front profiles but lack subordinate line emission. Profiles from groups 11 and 12 seem to be less exaggerated versions of the rarer profiles in group 52, which also appear on the ribbon front with broader widths, deeper central reversals and larger subordinate line emissions. In figure \ref{simp4}, we have plotted four flares during their impulsive phase. Panel A contains the upper ribbon of a two-ribbon flare and has profiles assigned to groups 11 and 12 at three different positions, each of which are following the direction of the ribbon, upwards for the highest point and downwards for the other two points. Similar behavior can be noted in panels B, C and D. In each case profiles from groups 11 and 12 follow the leading edge of the ribbon. Once the ribbon passes, profiles from groups 4 and 5 appear over the heated regions where the NUV emission is enhanced. When the ribbon starts to progress more slowly, the single peaked profiles from group 0 take their place, \textcolor{blue}{as can be seen in the online movies}.

\begin{figure}[tb] 
\centering
\includegraphics[trim={.8cm 1cm .8cm 2cm}, clip, width=0.5\textwidth]{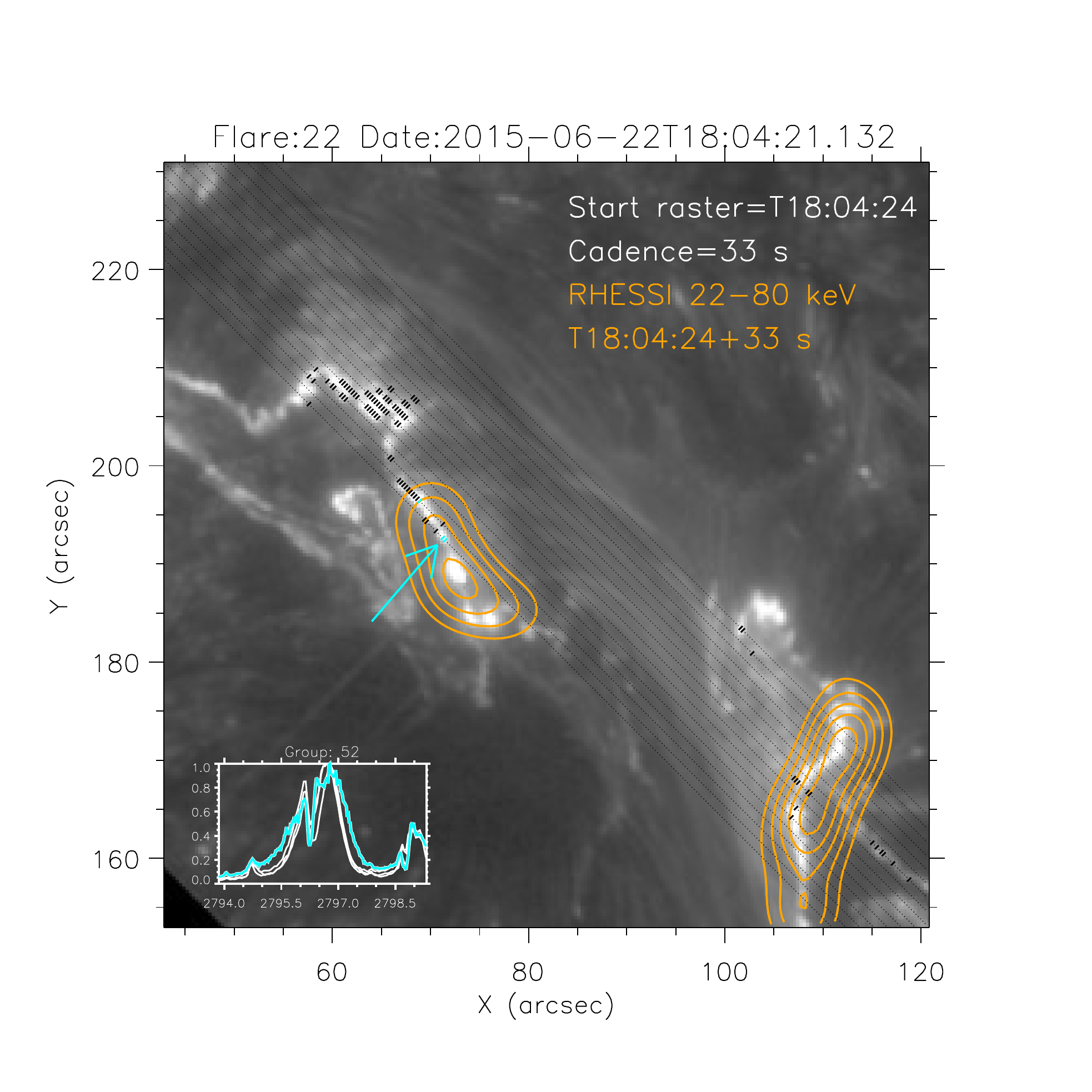}
\caption{IRIS 1400 SJI with slanted black lines marking the positions of the raster steps. The time of the first slit position is shown in the top right corner along with the rasters cadence. The RHESSI hard X-ray contours for levels [.20, .35, .50, .65, .80, .90] appear in orange, with the noise level of the image starting at .15. An insert in the bottom left hand corner shows the 3 spectra assigned to centroid 52. The cyan and black markers can be seen to follow the leading edge of the ribbon front in the \textcolor{blue}{online movies}. For clarity, only the ribbon-front profiles are shown.}
\label{RHESSI2}
\end{figure}

\begin{figure}[tb] 
\centering
\includegraphics[bb=22 45 555 790, clip, width=0.45\textwidth]{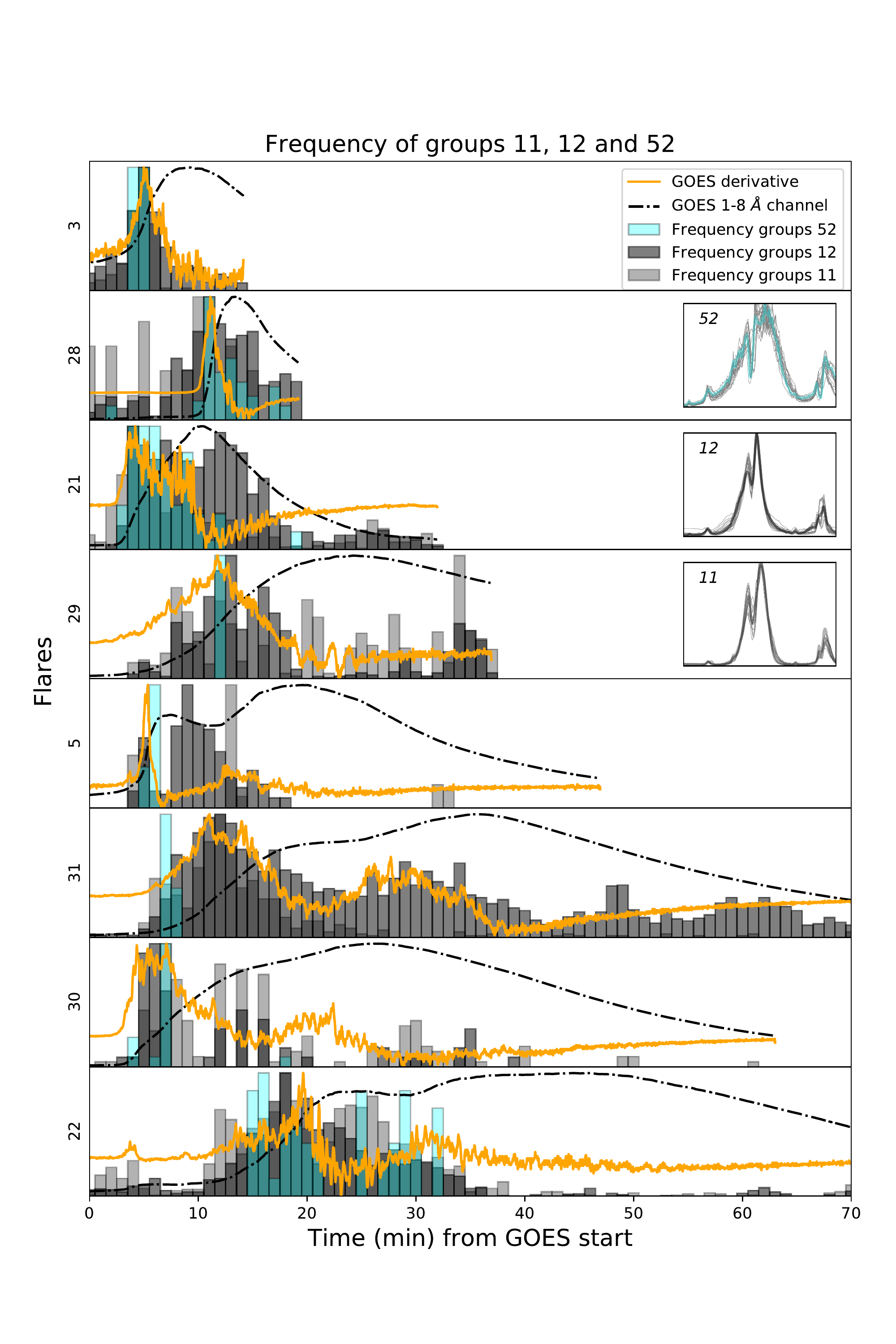}
\caption{Profile counts for groups 11, 12 and 52 plotted over the GOES $\text{1-8}~ \text{\AA}$ light curve and derivative. Subplots of the three groups have been included in flare panels 28, 21 and 29 with 20 of the profiles from that flare plotted in grey over their corresponding centroids. The histograms are normalized separately for each flare group. The ribbon-front IRIS profiles follow the GOES derivative and appear jointly in every flare observation.}
\label{X_ray_freq}
\end{figure}

Profiles belonging to group 52 have been observed by \cite{Lucia} (flare 28 of our list) to occur co-spatially and temporally with hard X-ray emission in the RHESSI 32-100 keV energy band. We found supporting evidence for the alignment of these profiles with hard X-ray signatures in the M6.5 flare observed both by IRIS and RHESSI on June 22, 2015  (flare 22 of our list). As seen in Figure \ref{RHESSI2}, one RHESSI hard X-ray source coincides with the cyan markers which indicate the positions of profiles from group 52. The contours were drawn using the "clean" reconstruction algorithm from a time integration equal to the cadence of the IRIS raster. The roll angle of RHESSI was not modified. The slanted black lines show the locations of the IRIS raster. Unfortunately, RHESSI was in earth's shadow during the impulsive phase of the flare, therefore we cannot be sure if the upper part of the ribbon had hard X-ray signatures previously, which may explain the black-colored profiles in that location and also the few cyan profiles seen in the movie at earlier times. Black ribbon-front profiles could still be related to X-ray emission if the X-ray emission is at least 10 times fainter than that of the main source, making them invisible due to RHESSI's limited dynamic range. In the \textcolor{blue}{online movies}, profiles from group 52 can be seen at the front of the fast moving bottom ribbon between times  T17:53:37-T17:57:00, but not exactly at the time shown in the figure. This could have several explanations: The profiles may have coincided with hard  X-rays at these times, but it cannot be verified due to RHESSI's orbit. The hard X-rays at 18:04 could be too weak to trigger the cyan profiles, or alternatively, hard X-rays do not necessarily trigger cyan-type profiles. They could also be hidden in this observation: post flare loops with large amounts of downflowing material can be seen in the IRIS movie and faintly in this figure. The viewing angle means that IRIS observes the lower ribbon through the loops. This may explain why instead of profiles from group  52 lining the lower ribbon, we see triangular profiles with large redshifts. In summary, cyan profiles seem to occur near hard X-ray contours, but there are exceptions and a future statistical study is desirable.

\begin{figure}[tb] 
\centering
\includegraphics[width=0.45\textwidth]{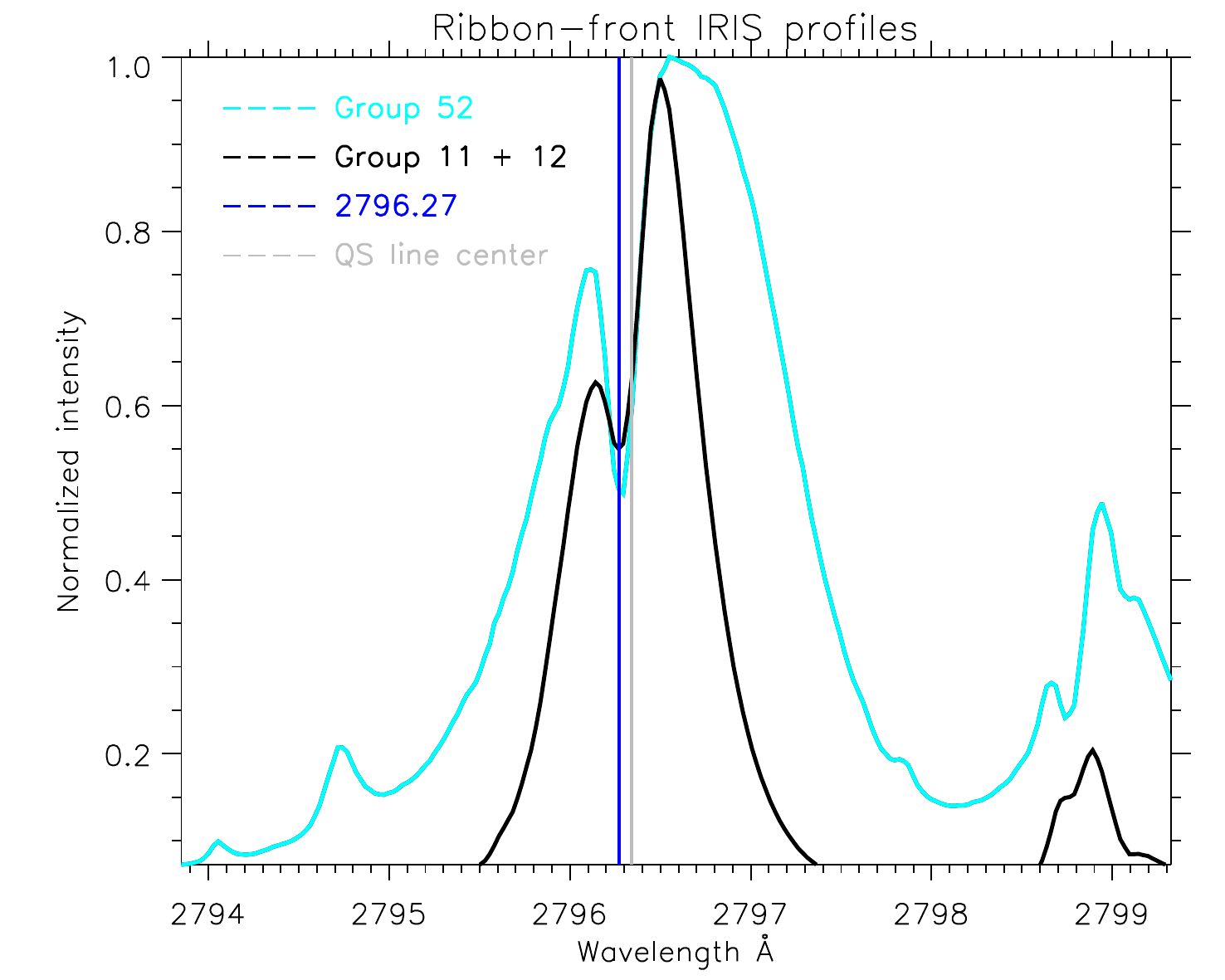}
\caption{Average profiles of groups 52 and (11+12) taken over all 33 flares, with non-thermal velocities of 95 and 54 km/s respectively. The position of the characteristic blue shifted reversal is indicated with a vertical blue line at  $2796.27$ and has a standard deviation of $0.04 ~\text{\AA}$.}
\label{X-ray1}
\end{figure}

In Figure~\ref{X_ray_freq} we have plotted the frequency of occurrence of the three ribbon-front IRIS profiles for 8 flares. The GOES curve and derivative in the 1-8 $\text{\AA}$ channel have been included for each flare to outline the impulsive phase. The three panels of flares 28, 21 and 29 contain the ribbon centroids as well as 20 overplotted profiles from the flare in that panel. Profiles belonging to group 52 occur during the impulsive phase and cluster around the maximum of the GOES derivative, in contrast to groups 11 and 12 which are more dispersed in relation to the GOES derivative, and may be the result of lower energy electron bombardment appearing to follow soft X-ray signatures. Profiles assigned to group 52 have FWHMs of $\sim1.5 ~\text{\AA}$ and characteristic blue shifted reversals at $2796.27 \pm  0.04 ~\text{\AA}$, whose wavelength calibration we have verified with quiet Sun profiles that were centered at 2796.34 \AA. Furthermore, these profiles may have large non-thermal contributions to their line width. Since IRIS has a negligible instrumental broadening, the non-thermal velocities can be calculated using the formula from \cite{NTB} given by
\begin{equation}
\xi_{nth} = \left[ \left(\frac{\text{FWHM}}{\lambda_0} \frac{c}{2\sqrt{2\text{log}2}}\right)^2 - \frac{2\text{k}_{\text{B}}\text{T}}{\text{m}}\right]^{1/2},
\end{equation}
where the second component is the squared doppler velocity and the line formation temperature T, was taken as $10^{4.4}$ K. Assuming the profiles are resolved, we find an upper limit of 95 km/s for the non-thermal velocities neglecting both pressure and opacity broadening. The velocities would be diminished if the profiles were unresolved. Figure~\ref{X-ray1} shows the typical line shapes of ribbon-front IRIS profiles. The cyan profile represents the average profile from group 52 that appears in 42\% of our flares. It is possible that IRIS misses the hard X-ray locations for some flares. The black profile is an average of every profile belonging to group 11 and 12 and occurs jointly in 100\% of the observations. Note the striking similarity between the two averaged profiles. Both have central reversals at precisely the same wavelength, and emissions from both the strong far red wing and its partner line forming a dimple. This leads us to believe that the ribbon profiles place a real physical constraint on the non-thermal velocity fields generated by the electron beam.

\subsection{Line ratios}
The intensity ratios of the \ion{Mg}{2} h\&k lines can be used as a diagnostic for the optical depth. A ratio of 2:1 of k:h indicates that the lines are formed in optically thin conditions, and a ratio of 1:1 indicates optical thickness. A derivation of these ratios is given in the Appendix.

In addition to the h\&k lines, there exists a companion of triplet lines from transitions between the $3\text{p}^2\text{P}$ and $3\text{d}^2\text{D}$ states. These subordinate lines are located on the blue and red wings of the k-line and have vacuum wavelengths of 2791.60, 2798.75 and 2798.82 \AA. The oscillation strength of the blue subordinate line (not visible in our window) is twice as weak as the far red line. From here on out, the strongest subordinate line at 2798.82 $\text{\AA}$ will be denoted by $s$. 

Early observations by \cite{firstkh} using NASA's Orbiting Solar Observatory recorded k/h ratios taken before and during a flare to be in the range 0.9-1.5. Quiet Sun ratios were measured by \cite{kh2} in the range 1.14-1.46, while more recent high resolution IRIS observations by \cite{Kerr} found time averaged quiet Sun ratios of $1.204\pm0.010$ and flaring ratios in the range 1.07-1.19. The SHK algorithm allows us to perform a large scale multi-flare study of the k/h ratios using the groups to partition flaring and non-flaring profiles. Additionally, we monitor the trend of the k/$s$ ratios. 

IRIS's sensitivity slowly degrades over time, therefore the spectral data must be recalibrated for each observation. Before calculating the line ratios we took into account the change in effective area and converted the measured intensities $I_m(\text{DN/s})$ into physical units $I_{phys}(\text{erg}~\text{s}^{-1}\text{cm}^{-2}\text{sr}^{-1}\text{\AA}^{-1})$, see \cite{Lucia2} for details. The ratios were then calculated by dividing the integrated intensities of the h\&k lines taken over a $2~\text{\AA}$ window. For a window exceeding  $2~\text{\AA}$, the asymmetries in the wings, which contain a number of blended lines, result in inaccuracies. 

\begin{figure}[tb] 
\centering
\includegraphics[trim={0cm .5cm 0cm 1cm}, clip, width=0.5\textwidth]{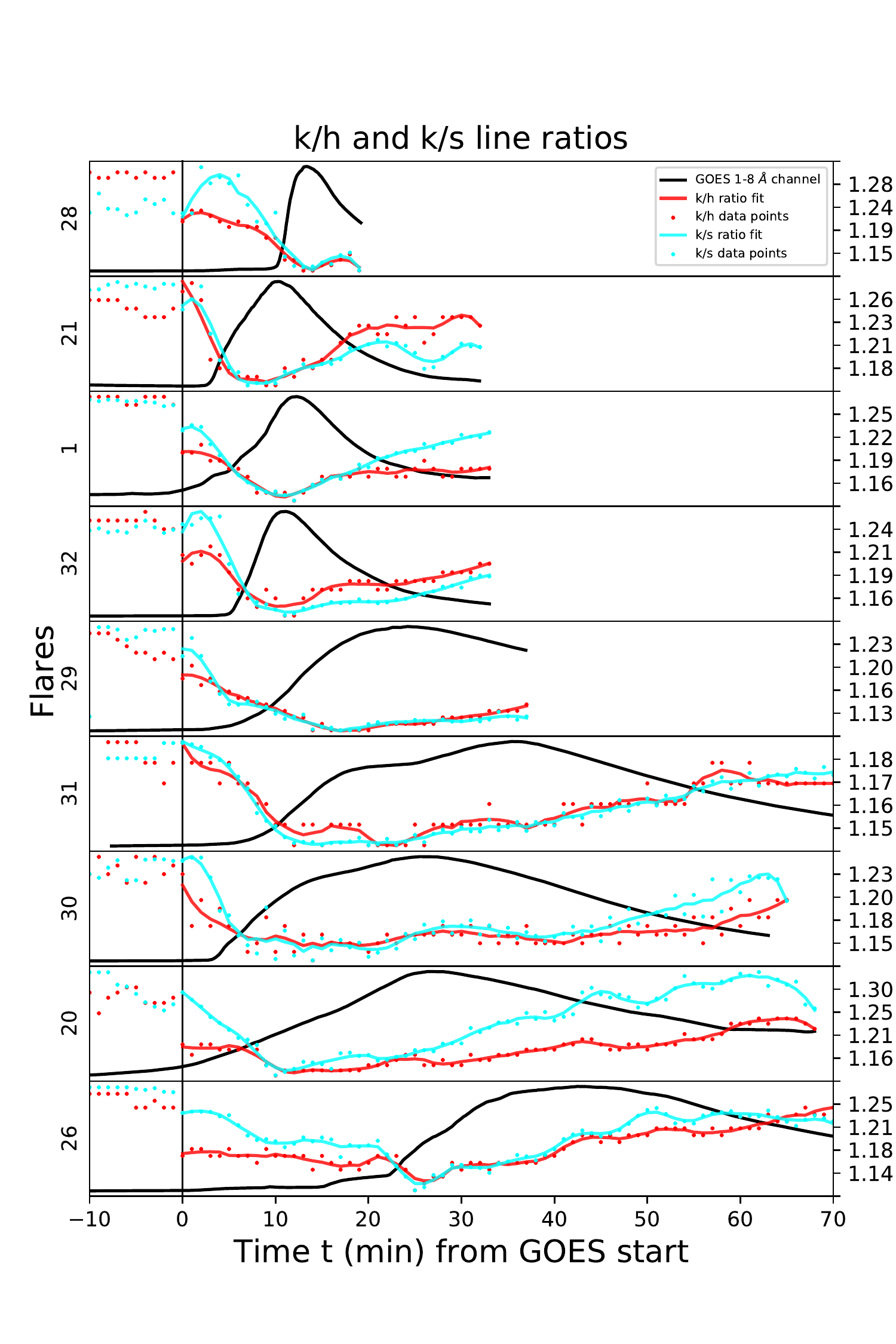}
\caption{Plot of line ratios for several flares, with the GOES curves given by solid black-lines. The raw data points for the k/h ratio (red) and k/$s$ ratio (cyan) are averages of all measured values within the specified minute window. The k/h ratios can be inferred from the scale on the right of each panel. A vertical black-line separates ratios calculated using non-flaring profiles for times $t<0$ and ratios calculated from flaring profiles for times $t\geq0$. The solid red and cyan lines were generated by fitting a third order polynomial to the raw flaring data in a running widow of width 9. Both the k/h and k/$s$ ratios decrease during each flare, indicating an increase in opacity and enhanced subordinate line emission respectively.}
\label{ACDC}
\end{figure}
Figure \ref{ACDC} shows the evolution of both the k/h and k/$s$ line ratios. For times $t\textless0$, the ratios of profiles belonging to quiet Sun groups were used, while profiles from flaring groups were used for times $t\geq0$. This division is marked with a black vertical line across all panels.  The raw data appears as points with each point representing the average measurements of ratios occurring within the specified  minute window. In order to clearly show the trends, the data points of the k/h and $\text{k}/s$ ratios for $t\geq0$ have been fitted with a third order polynomial applied over a running widow of width 9. The numerical scales of the k/h-line ratios for each flare are given on the right-hand side of each panel. It is clear that both the k/h and k/s ratios decrease during each flare, indicating an increase in opacity (due to enhanced electron densities) and enhanced subordinate line emission respectively.

\begin{figure}[tb] 
\centering
\includegraphics[trim={0cm 0cm 0cm .7cm}, clip, width=0.5\textwidth]{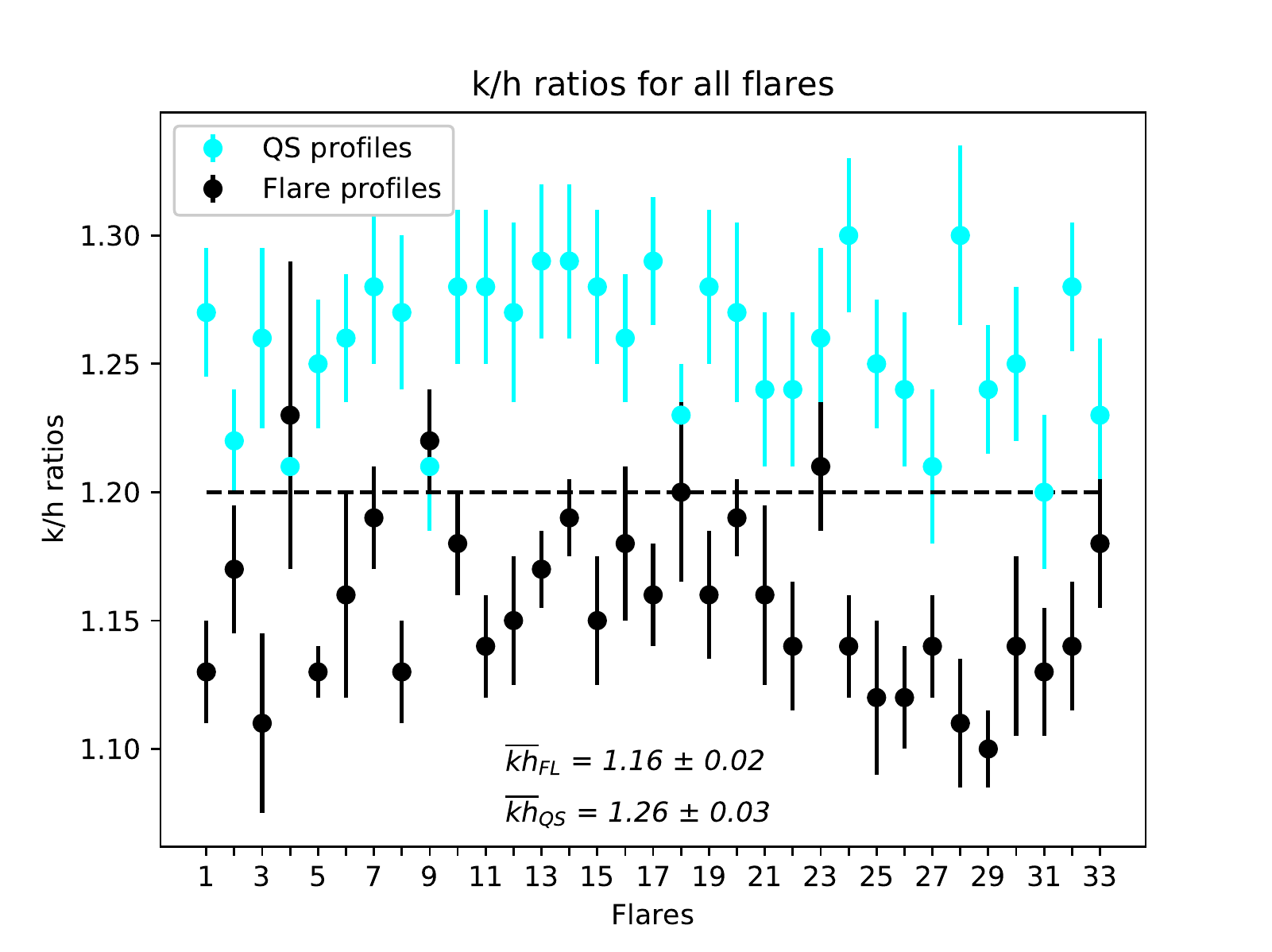}
\caption{Average k/h ratios for QS and flaring regions using the representative groups (39, 43, 38, 41, 42, 44) and (4, 5, 8, 11, 12, 52). The horizontal dashed line separates flaring and non-flaring k/h ratios based on the current literature. Flare 4 and 9 had very few profiles assigned to the chosen centroids.}
\label{kh2}
\end{figure}

In Figure \ref {kh2}, we partitioned the ratios into flaring and non-flaring ratios based on a few representative quiet Sun and flaring groups. The average k/h ratios for flaring profiles across all 33 flares were found to be $1.16\pm0.02$ in comparison to the quiet Sun values of $1.26\pm0.03$. The large variance in the ratios is a natural consequence of partitioning the profiles into groups. The results agree with the current ratio calculations of \cite{Kerr} based on flare number 15 of our list.
 
\begin{figure}[tb] 
\centering
\includegraphics[trim={0cm 0cm 0cm 0.7cm}, clip, width=0.5\textwidth]{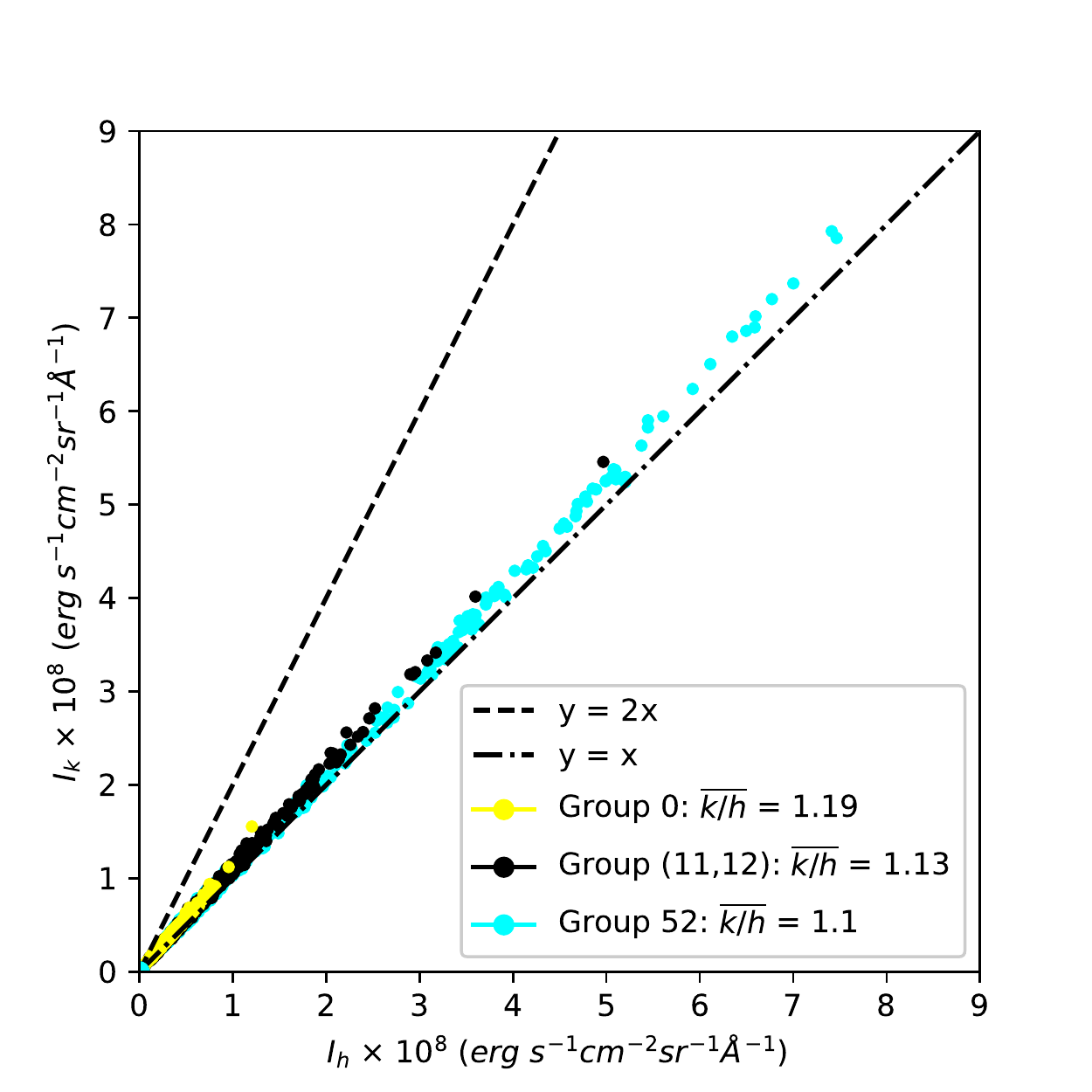}
\caption{k/h ratios for groups 0, (11+12) and 52 for all 33 flares with h\&k intensities given in absolute units. The optically thick and thin ratios are represented by the dashed lines $y=x$ and $y=2x$. Only the high intensity portion of the range $10^5-10^8$ has been plotted. Higher intensity profiles generally occur closer to the optically thick ratio, implying that during a flare, the h\&k lines remain optically thick.}
\label{thick_thin}
\end{figure}

The intensities in absolute units span a range of $10^5-10^8$ $(\text{erg}~\text{s}^{-1}\text{cm}^{-2}\text{sr}^{-1}\text{\AA}^{-1})$. In Figure \ref{thick_thin}, we plotted the high intensity portion of ratios of all 33 flares, and color coded ratios from profiles belonging to group 0, (11+12) and 52. The ratios remain far away from the optically thin $y=2x$ line, and in general, higher intensity profiles occur closer to flare maximum and correspond to larger opacities. These results are not surprising since both emission and opacity depend on electron density, which based on half-width measurements of high Balmer lines with $\tau_0<1$ can reach values of $n_e\sim4\times10^{13}cm^{-3}$ at flare maximum, and vary over two orders of magnitude during a flare \citep{edensity}. We conclude that the \ion{Mg}{2} lines not only remain optically thick during a flare, but appear closer to the 1:1 ratio than in the quiet Sun.

\section{Discussion}
\subsection{Typical flare profiles}

In this section, we review the common features of flare profiles and compare these features to that of the quiet Sun. We refer to the known mechanisms of quit Sun profile formation, and discuss the likely physical process responsible for the observed differences.

The quiet Sun profiles are observed almost completely at their $\tau=1$ value, and have a formation height that extends over the entire chromosphere, with the left (k1v) and right (k1r) minimums being formed in the lower chromosphere, left (k2v) and right (k2r) peaks in the middle chromosphere and line core photons coming from the upper chromosphere, just below the transition region. The wings contain contributions from the photosphere, and because they are formed under LTE conditions at $\tau=1$, they follow the source function and consequently the lower atmospheric temperature structure. This results in raised wings as one samples lower and lower heights of the photosphere. The wings are most noticeable in profiles belonging to groups 44 and 45 in Figure \ref{centroids}. The raised wings become "flattened" for emissions over sunspots or during flares. For sunspots, the temperature is about $\sim1500$ K lower than the quiet Sun due to the stifling of convective energy by large kilogauss magnetic fields \citep[e.g.][]{Sunspot}, which results in a diminished source function. For flares the flattening could possibly be explained by a downwards condensation layer that forms an optically thick barrier between the photosphere and the rest of the atmosphere \citep{Adam}. The resulting emergent intensities would then only contain contributions form the chromosphere.

In addition to the flattened wings, we found single-peaked profiles to be prevalent in every flare and discuss their potential origin here. Once again, under quiet Sun conditions the \ion{Mg}{2} h\&k line profiles have a characteristic central reversal. This reversal is common in strong chromospheric lines and is due to the source and Plank function decoupling at heights comparable to the core formation height. \cite{rev} found a frequency independent approximation of the source function that holds for upper chromospheric heights, given by
\begin{equation}
S = \frac{\int J_v\phi_v dv + \epsilon B_v(T_e)}{1+\epsilon},
\label{source}
\end{equation} 
with mean intensity $J_v$, absorption profile $\phi_v$, Planck function $B_v(T_e)$ and photon destruction probability $\epsilon$ approximated as the ratio of the collisional and spontaneous de-excitation coefficients $C/A$. At heights close to the core formation height, the source function decreases on account of $A$ being several orders of magnitude larger than $C$.

As explained by \cite{Paper1} and \cite{Decop}, the h\&k line core photons are produced in the thin upper chromosphere where large photon mean free paths and low photon destruction probabilities prevail. As such, the radiation field is horizontally smoothed and the Eddington approximation holds. Consequently, the source function along with the emergent intensity decreases with height, resulting in an observed central reversal. We find that during a flare, the central reversals are commonly seen in emission. Eq. \ref{source} demonstrates that an increase in electron density and temperature could result in a source function that continues to increase with height. A recent parameter study by \cite{Lucia} found that central emissions were indeed stimulated either by a large temperature spike in the upper chromosphere or an increase in electron density at the same location. In both cases the coupling of the source and Planck function persists to heights above the core formation height, allowing the entire profile to form under LTE. The study demonstrated that unresolved up- and downflows can also lead to single peaked profiles formed under non-LTE conditions. In this case, the complex velocity fields containing both condensation and evaporation patterns can produce photons that fill the central reversal.  It is unclear which of the three mechanisms are responsible for the single peaked flare profiles, which may result from an interplay between all three processes.

The ribbon would provide the most complementary conditions for single peak production, with enhanced electron densities, temperatures and doppler velocities, and therefore their prevalence in flares may be explained. 

\subsection{Profiles at leading edges of flare ribbons}
Recent observations of the ribbon front have shown that NUV spectra resulting from the non-thermal electron beam may differ from other spectra within the field of view. \cite{Negative_Flare_Front} observed a negative flare front in \ion{He}{1} 10830 $\text{\AA}$ using the Goode Solar Telescope at Big Bear Solar Observatory. The ribbon front produced the broadest \ion{Mg}{2} line profiles in the field of view, with a FWHM of 1 \AA. \cite{Blue} observed the leading edge of a C-class flare kernel on November 11, 2014 with IRIS. They observed \ion{Mg}{2} profiles with intensity enhancements in the blue wing and smaller blue (h2v) peaks in comparison to the red (h2r) peaks of the h-line. The apparent blueshifts lasted 9-48 s, with speeds of $10.1\pm 2.6$ km/s and were followed by strong redshifts up to 51 km/s. They proposed a simple model where non-thermal energetic electrons would heat a deep region of the atmosphere which would then expand and carry cool chromospheric-temperature plasma into the corona. This mechanism was verified by a simple non-LTE cloud model \citep{Cloud}, where the emission of the rising cool plasma explained the blue wing enhancement, and the peak asymmetries were naturally reproduced on account of the cool upwards moving material shifting the maximum opacity of the line into the blue. The peak asymmetries and blue wing enhancements were not observed in lines such as  \ion{Ca}{2} K, \ion{Ca}{2} 8542 $\text{\AA}$ and $\text{H}\alpha$, which have significantly smaller optical depths than the \ion{Mg}{2} lines. Additionally, using a non-LTE simulation with partial redistribution taken into consideration for the line cores, \cite{Lucia} generated synthetic blue shifted h\&k lines with the desired blue and red peak asymmetries by combining two spectral profiles produced by downflows at different chromospheric heights.

We find that profiles located at the ribbon front are often the broadest profiles in the field of view, only surpassed by even broader profiles due to filament eruptions (group 10) or large downflows at the end of the flares (group 8). Large broadenings may indicate large non-thermal velocities and thus turbulence. We therefore conclude that if the profiles are resolved, the turbulence is highest at the leading edge of the flare ribbon. 

Profiles from the combination of groups 11 and 12 occur in every flare, although tenuously and with extremely low triplet emissions in the succession of M-class flares on October 26, 2014. However, the slit of IRIS was positioned poorly in relation to the active region's major ribbon activity. On this day, flare 9 produced a profile assigned to group 11 with precise width and peak asymmetries, but with no triplet emission. The profile seems to be caused by upflowing material from a clearly visible erupting jet. 

We conclude that the ribbon-front IRIS profiles have three possible origins: 1) They are generated by the superposition of unresolved downflows at different chromospheric heights, as demonstrated by \cite{Lucia}. 2) They come about due to enhanced turbulence at the leading edge of the ribbon front, with triplet line emission due to the heating of the lower chromosphere, in line with the quiet sun triplet modeling of \cite{triplet_diagnostics}. 3) They are generated by rising cool chromospheric-temperature material as discussed by \cite{Blue}. The last explanation seems unlikely since flare 9 and the cloud model of their study generated similar profiles to our ribbon-front IRIS profiles but without the subordinate emission. This makes it clear that there should be a distinction between profiles generated from upflowing material and profiles generated from non-thermal electron bombardment. This distinction is based on the engagement of the two red wing triplet lines.

\section{Conclusions and outlook}
We have shown that clustering based on machine learning is a very suitable tool to systematically identify and analyze spectra in a multitude of flares. Our main results can be summarized as follows:
\begin{itemize}
\item Typical \ion{Mg}{2} h\&k flare profiles consisting of a single peak and broad wings exist and appear in all flares. They are located over heated regions where the NUV emission is enhanced. The subordinate lines appear in emission.
\item Profiles at the leading edges of flare ribbons appear to follow X-ray signatures and also have typical shapes. They are generally the broadest profiles in the field of view and contain a blue-shifted reversal at 2796.27 \AA.
\item The k/h-line ratios for flaring and non-flaring profiles are well separated, with values of $1.16\pm0.02$ and $1.26\pm0.03$ respectively. During a flare, this ratio decreases with higher intensity profiles appearing closer to the optically thick 1:1 ratio.
\end{itemize}

We may explain the prevalence of the single peaked profiles either through enhanced electron densities, temperatures, or unresolved up and downflows, all of which are expected to occur in the wake of flare ribbons, as a consequence of the high electron deposition rates from the corona (see section 4.1). The unusual shape of the profiles at the leading edge of flare ribbons could be generated either by unresolved downflows or enhanced turbulence. In the case of unresolved downflows, a dip at a surprisingly constant wavelength may indicate similar downflows for all flares, and was also shown to occur in simulations. In case the profiles are resolved, the enhanced turbulence would be greatest at the leading edge of the flare ribbon, which also is expected from flare models.

Furthermore, we suggest that the ribbon-front IRIS profiles along with the single peaked profiles are universal flare indicators, with the single peaked profiles linked to the generic heating and increased electron density of the chromosphere during a flare, and the ribbon-front IRIS profiles being the NUV counterpart to the magnetic reconnection event.\\ 
 
Extending the SHK analysis to include additional lines such as \ion{Si}{4}, \ion{C}{2} and \ion{O}{4} will allow us to analyze the atmospheric conditions at different heights and in more detail, leading to a better understanding of the flaring atmosphere. Rare profiles such as those belonging to group 16 should be understood more thoroughly using forward modeling with realistic flaring atmospheres and non-LTE radiative transfer codes. SHK may also be an ideal and reliable candidate for detecting downflows over the entire IRIS database.

Additionally, we plan on performing an exhaustive study of co-aligned IRIS and RHESSI observations to provide further evidence of the spatial and temporal relationship between hard X-ray emission and profiles from group 52.\\
  
 We would like to thank the Swiss National Science Foundation for funding this research under grant number 407540\_167158, as well as LMSAL and NASA for allowing us to download all the IRIS data from their servers. IRIS is a NASA small explorer mission developed and operated by LMSAL with mission operations executed at NASA Ames Research center and major contributions to downlink communications funded by ESA and the Norwegian Space Centre. SK has been funded through NASA contract NAS 5-98033.

\bibliographystyle{apj}
\bibliography{journals,references}

\appendix
\section{Derivation of the k/h ratio}

The collision rate between the h\&k levels and the ground state is much smaller that the spontaneous radiative de-excitation rate \citep{Paper1}. Consequently, the line strengths are only loosely coupled and we can expect the emergent intensities of the two spectral lines to differ from one another appreciably. The flux $F(\lambda_{ij})$ recorded on earth is given by
\begin{equation}
F(\lambda_{ij}) = P\frac{1}{4\pi R}\int n_jA_{ji}~dV,
\end{equation}
where $\lambda_{ij}$ characterizes the wavelength of photons from the transition $j\to i$, $R=1.5\times10^{13}~\text{cm}$, $n_j$ is the atomic population of the h or k level, $A_{ji}$ is the Einstein coefficient for spontaneous emission from the upper state j to the lower state i, $V$ is the volume and $P$ is the photon escape probability. Following the comprehensive review of photon escape probabilities by \citep{escape}, the mono-directional, frequency averaged single-flight escape probability for a doppler profile with constant emission along the line of sight is given by
\begin{equation}
P(D,\overline{k},\tau_0) = \frac{1}{\tau_0\sqrt{\pi}}\int^\infty_{-\infty} [1-\text{exp}(-\tau_0 e^{-x^2})] \text{d}x,
\label{Pescape}
\end{equation}
where $D$ means we have averaged over a doppler profile, $\overline{k}$ is the direction of emission, $\tau_0$ is the optical depth at the center of the profile and $x$ is a dimensionless frequency variable. This equation stems form the assumption that the escape probability depends on the optical depth in the same manner as the one-dimensional equation of transfer, namely $\text{exp}(-\tau_v)$.
For a detailed numeric calculation of Eq. \ref{Pescape} see \cite{numeric}. The asymptotic behavior at $\tau_0\to0$ and $\tau_0\to\infty$ is relatively simple to calculate. For the optically transparent case, the integral simplifies to a Gaussian integral after expanding the exponential in a power series and taking the limit $\tau_0\to0$, giving $P(D,\overline{k},\tau_0)=1$, while for the optically opaque case, the escape probability can be written as a convergent series which in the limit $\tau_0\to\infty$ gives $P(D,\overline{k},\tau_0)= 2(\text{log}\tau_0)^{1/2}/\sqrt{\pi}\tau_0$. 

In general, the flux ratio between two lines reduces to a product of the ratios of the photon escape probabilities, statistical weights of the lower level, element abundance, ionization fraction and collision strengths given by 

\begin{equation}
\Omega_{ij} = \frac{8\pi}{\sqrt{3}}\frac{I_H}{\Delta E_{ij}}g\omega_if_{ij},
\end{equation}
where $I_H$ is the ionization energy of hydrogen, g a Gaunt factor to correct for quantum mechanical effects, $\Delta E_{ij}$ the threshold energy required for the transition and $f_{ij}$ the levels oscillation strength. Because we are analyzing two resonant lines from the same element and ionization state, the k/h ratio will reduce to the ratio of each levels oscillation strength $(f_{ik}=2f_{ih})$ multiplied by the ratio of each core wavelengths escape probability
\begin{equation}
\frac{F(\lambda_{ik})}{F_(\lambda_{ih})} = \frac{f_{ik}}{f_{ih}}\frac{P(D,\overline{k},\tau_k)}{P(D,\overline{k},\tau_h)}.
\end{equation}
The h\&k lines can therefore be used as an opacity diagnostic, with the optically thin case correspond to a flux ratio of 2:1 and the optically thick case correspond to a flux ratio of 1:1.

\begin{table}[h]
\caption{Flare List} \centering \small \label{Flares_List}
\begin{tabular}{llllllll}
\toprule\toprule
Flare &  Class  &  ~~~Date and time &  ~~~~~~~Observation mode   & CAD  &~ FOV   &FOV center & ~~OBSID \\ 
& &  when raster started& &(sec)&($\text{arcsec}^2$)&~~(arcsec)&\\ \midrule
1&  M1.0&  2014-06-12T18:44& Medium coarse 8-step raster&  21.34& $14\times60$& (-670,-306)& 3863605329 \\ 
2& M1.0 &2014-10-26T18:52 &Large sit-and-stare& 16.20& $0.3\times120$& (648,-287)& 3864111353 \\ 
3& M1.0 &2014-11-07T09:37& Large coarse 16-step raster& 23.34 & $30\times120$& (-646,224)&  3860602088 \\
4& M1.0&  2014-10-26T15:31& Large sit-and-stare& 5.36& $0.3\times120$& (598,-307)& 3880106953 \\
5& M1.1& 2014-09-06T11:23& Large sit-and-stare& 9& $0.3\times120$& (-709,-298)& 3820259253 \\
6 &  M1.1&  2015-08-21T16:01& Medium dense 32-step raster& 102& $10\times60$& (-467,-336)& 3660104044\\
7& M1.3& 2014-02-02T21:08 & Very large coarse 64-step raster& 2051& $126\times175$& (7,-123)& 3880012095 \\
8& M1.3& 2014-06-12T11:09& Medium coarse 8-step raster& 21& $14\times60$& (-723,-303)& 3863605329 \\
9& M1.3& 2014-10-26T18:52& Large sit-and-stare& 16& $0.3\times120$& (648,-287)& 3864111353 \\
10& M1.4& 2015-03-12T05:45& Large sit-and-stare& 5& $0.3\times120$& (-185,-190)& 3860107053 \\
11& M1.5& 2014-02-04T15:30& Large dense 64-step raster& 1084& $21\times120$& (245,-100)& 3880010190\\
12& M1.5& 2014-08-01T17:20& Large dense 64-step raster& 2036& $21\times120$& (-200,-230)&3800013190\\
13& M1.6&2015-03-12T05:45&sit-and-stare&5&$0.3\times120$&(-185,-190)&3860107053\\
14& M1.8&2014-02-11T16:34&Very large dense 64-step raster&2049&$21\times175$&(-197,-123)&3880012191\\
15& M1.8&2014-02-12T21:50&Large coarse 8-step raster&42&$14\times120$&(140,-90)&3860257280\\
16& M1.8&2015-03-11T04:46&Large coarse 8-step raster&75&$14\times120$&(-430,-194)&3860259280\\
17& M2.3&2014-11-09T15:17&Large coarse 4-step raster&37&$6\times120$&(-217,-205)&3860258971\\
18& M2.4&2014-10-26T18:52&Large sit-and-stare&16&$0.3\times120$&(648,-286)&3864111353\\
19& M2.9&2015-08-27T05:37&Large coarse 8-step raster&24&$14\times120$&(24,708)&3860605380\\
20& M3.4&2014-10-27T20:56&Large sit-and-stare&16&$0.3\times120$&(16,779)&3864111353\\
21& M3.9&2014-06-11T18:19&Medium coarse 8-step raster&21&$14\times60$&(-781,-306)&3863605329\\
22& M6.5&2015-06-22T17:00&Large sparse 16-step raster&33&$15\times120$&(72,192)&3660100039\\
23& M6.6&2014-10-27T20:56&Large sit-and-stare&16&$0.3\times120$&(779,-271)&3864111353\\
24& M7.0&2014-04-18T12:33&Large sit-and-stare&9&$0.3\times120$&(568,-230)&3820259153\\
25& M7.1&2014-10-26T18:52&Large sit-and-stare&16&$0.3\times120$&(648,-287)&3864111353\\
26& M8.7&2014-10-21T18:10&Large sit-and-stare&16&$0.3\times120$&(-359,-316)&3860261353\\
27& X1.0&2014-10-25T14:58&Large sit-and-stare&5&$0.3\times120$&(408,-319)&3880106953\\
28& X1.0&2014-03-29T14:09&Very large coarse 8-step raster&72&$14\times175$&(490,282)&3860258481\\
29& X1.6&2014-09-10T11:28&Large sit-and-stare&9&$0.3\times120$&(-137,125)&3860259453\\
30& X1.6&2014-10-22T08:18&Very large coarse 8-step raster&131&$14\times175$&(-292,-303)&3860261381\\
31& X2.0&2014-10-27T14:04&Large coarse 8-step raster&26&$14\times120$&(727,-299)&3860354980\\
32& X2.1&2015-03-11T15:19&Large coarse 4-step raster&16&$6\times120$&(-353,-197)&3860107071\\
33& X3.1&2014-10-24T20:52&Large sit-and-stare&16&$0.3\times120$&(264,-302)&3860111353\\
 \bottomrule
\end{tabular}
\end{table}

\begin{table}[h]
\caption{Non-flaring observations List} \centering \small \label{Non_flares_List}
\begin{tabular}{lllllllll}
\toprule\toprule
 &  Target &~~~Date and time  &  ~~~~~~~Observation mode   & CAD  &~ FOV   & FOV center & ~~OBSID \\ 
& &when raster started & &(sec)&($\text{arcsec}^2$)&~~(arcsec)&\\ \midrule
A&Quiet Sun&2014-06-07T07:29& Large sit-and-stare& 17& $0.3\times120$& (128,-574)& 3820011653 \\ 
B&Supersonic downflow&2014-09-10T21:29& Very large sit-and-stare& 5& $0.3\times175$& (-71,111)& 3800507454 \\ 
C&Brightenings near AR&2014-10-01T11:25&Large coarse 2-step rast. & 18& $2\times120$& (-441,-125)& 3860359362 \\ 
D&Brightenings near AR&2014-10-23T21:39&Large sit-and-stare&16& $0.3\times120$& (48,-301)& 3860261353\\ 
E&Pores and flux emergence&2014-11-02T18:48&Large coarse 4-step rast. &37& $6\times120$& (-206,-327)& 3860258971\\ 
F&Flux emergence region&2014-11-09T02:16&Large sit-and-stare&10& $0.3\times120$& (-292,177)& 3860009153\\ 
G&Small brightenings in spot&2014-11-19T14:08&Very large sit-and-stare&10& $0.3\times175$& (-139,-298)& 3860259254\\ 
H&Jets/tiny flare&2014-11-21T05:09&Very large sit-and-stare&10& $0.3\times175$& (186,-284)& 3860259254\\ 
I&Sunspot with light bridge&2014-12-06T06:03&Large sit-and-stare&5& $0.3\times120$& (725,-342)&3860256053\\ 
 \bottomrule
\end{tabular}
\end{table}

\begin{figure}[h] 
\centering
\includegraphics[trim={.5cm .8cm .5cm .5cm}, clip, width=1\textwidth]{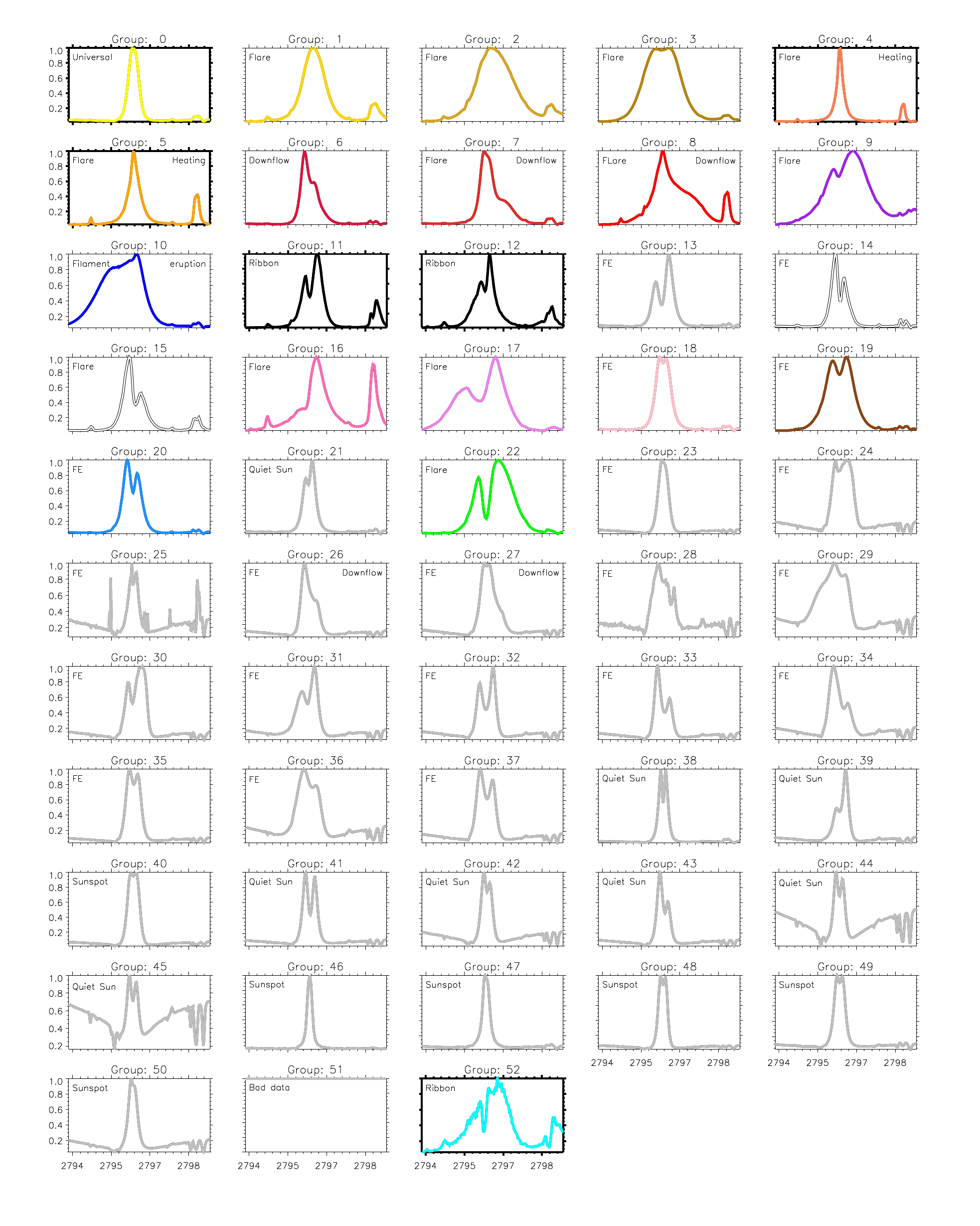}
\caption{Centroid colors as seen in the online movies.}
\label{centroids_for_movies}
\end{figure}

\end{document}